\documentclass[twocolumn,nofootinbib,amsmath,amssymb,prl]{revtex4}

\usepackage{graphicx}
\usepackage{dcolumn}
\usepackage{bm}
\usepackage{color}
\usepackage{cancel}
\usepackage{subfigure}

\def\max{{\rm max}}

\def\d2l{\frac{d^2l}{(2\pi)^2}}

\newcommand{\bea}{\begin{eqnarray}}
\newcommand{\eea}{\end{eqnarray}}
\newcommand{\be}{\begin{equation}}
\newcommand{\ee}{\end{equation}}
\newcommand{\ben}{\begin{eqnarray}}
\newcommand{\een}{\end{eqnarray}}

\begin{document}

\title{21-cm Lensing and the Cold Spot in the Cosmic Microwave Background}

\author{Ely D. Kovetz${}^{1}$ and Marc Kamionkowski$^{2}$}

\affiliation{$^1$Theory Group, Department of Physics and Texas
Cosmology Center,
The University of Texas, Austin, Texas 78712, USA}
\affiliation{$^2$Department of Physics and Astronomy, Johns Hopkins University, Baltimore, MD 21210, USA}


\begin{abstract}
An extremely large void and a cosmic texture are two possible
explanations for the cold spot seen in the cosmic
microwave background. We investigate how well these two
hypotheses can be tested with weak lensing of 21-cm fluctuations
from the epoch of reionization (EOR) measured with the Square
Kilometer Array (SKA).  While the void explanation for the cold
spot can be tested with SKA, given enough observation time, the
texture scenario requires significantly prolonged observations,
at the highest frequencies that correspond to the EOR, over the
field of view containing the cold spot.
\end{abstract}


\maketitle

A number of hypotheses have been advanced to explain the
temperature decrement inside a circle of radius $5^{\circ}\mbox{-}10^{\circ}$ in the direction 
$(b,l)\sim(-57^{\circ},209^{\circ})$ (in Galactic coordinates)
in the cosmic microwave background (CMB), 
which is known as the WMAP cold spot.  It first drew attention when convolved with a spherical Mexican-hat wavelet filter of a corresponding radius, it was found to be in tension with $\Lambda {\rm CDM}$ at a $\gtrsim98\%$ C.L. \cite{Vielva}. Although using simpler filters exhibited a weaker tension \cite{ZhangHuterer}, this sky region has shown up as a significant candidate in several other CMB analyses looking for various particular signatures \cite{Feeney1,McEwen,Feeney2}.
While the simplest conjecture is that it is merely a statistical fluke, a possible hypothesis which we explore in this letter is that it results from an extremely large void in the direction of the cold spot \cite{InoueSilk, Holman, Cruz}.  In fact, a detection of an (under)dense region in the NRAO VLA Sky Survey in the vicinity of this direction was reported \cite{Rudnick} but then later disputed \cite{SmithHuterer}. Other attempts to observe such a void did not find any supporting evidence \cite{Granett,Bremer}. Another hypothesis \cite{ScienceTexture, Cruz} posits a cosmic texture \cite{Turok} as a source for the cold spot. 

Both a void and a texture would serve as sources of
gravitational lensing. Their CMB lensing signatures have
already been proposed in Ref.~\cite{DasSpergel}.  There it was
claimed that high-resolution CMB experiments like SPT or ACT
would be able to detect these signals provided that the
telescopes are aimed at the target for a sufficiently long
time, the texture requiring at least an order-of-magnitude more
exposure than the void.  However, a more recent analysis
\cite{BeNaSunny} (see also \cite{Masina}) disputed these claims, concluding that the
CMB lensing signatures of both the texture and the void would be
undetectable by high-resolution CMB experiments. 

Fluctuations in the 21-cm emission from neutral hydrogen during
the epoch of reionization (redshifts $7\lesssim z \lesssim 13$)
can be used in place of the CMB as a source for lensing by the
texture or void.  The 21-cm signal has two potential
advantages over the CMB for detecting lensing by local
structures:  (1) Fluctuations in
the 21-cm background are damped only at the baryonic Jeans mass,
which corresponds to a multipole $l\gtrsim10^6$; (2) by
observing at multiple frequencies we obtain many different
statistically independent 21-cm backgrounds to serve as sources
for the lens.  These can then be combined to reduce the noise in
the lensing reconstruction.

Lensing of the EoR 21-cm signal may be sought with the Square
Kilometer Array (SKA) \cite{ska}, a
next-generation low-frequency radio interferometer.  SKA will
scan the lower Galactic hemisphere and will map the intensity of
the 21-cm background from the EoR in direction of the WMAP cold
spot.  In this Letter we evaluate the prospects to detect the
lensing signal from the void and the texture postulated to
account for the CMB cold spot with these 21-cm maps.
We consider a method introduced in Ref.~\cite{KovKam} to detect a cluster via lensing reconstruction of 21-cm fluctuations.  We then evaluate the signal-to-noise with which the void or texture responsible for the WMAP cold spot can be detected with a given choice of experimental parameters for SKA.

The power spectrum for intensity fluctuations in the 21-cm 
signal from emission during the EoR on the scales relevant for
lensing reconstruction of local structure behaves
roughly as $l^2C_l \sim {\rm const} \equiv
2\pi\alpha(\nu,\Delta\nu)$
\cite{KovKam,Zaldarriaga}, as a function of
angular multipole $l$, where $\nu$ is the observation frequency,
and $\Delta\nu$ is the frequency bandwidth of the experiment.

The noise power spectrum of the interferometer is given by
\be
     l^2 C_l^n= \frac{(2\pi)^3 T_{\rm sys}^2(\nu) }{ \Delta \nu
     t_o  f_{\rm cover}^2} \left(\frac{ l}{ l_{\rm
     cover}(\nu)}\right)^2,
\ee
where $l_{\rm cover}(\nu)=2\pi D/\lambda$ is the maximum
multipole at frequency $\nu$ (corresponding to wavelength
$\lambda$) that can be measured with an array of dishes with
maximum baseline $D$
covering a total area $A_{\rm total}$ with a covering fraction
$f_{\rm cover}\equiv N_{\rm dish} A_{\rm dish} /A_{\rm total}$
in a frequency window $\Delta \nu$ with an observing time
$t_o$. The system temperature is given by $T_{\rm sys} 
\sim 180\left(\nu/180\,{\rm MHz}\right)^{-2.6}$~K.

Defining $\beta(\nu,\Delta\nu)\equiv(2\pi)^2 T_{\rm sys}^2(\nu)
\left[ l_{\rm cover}(\nu)^2 \Delta\nu \right]^{-1}$, we find that the
maximum multipole where the signal power
spectrum $C_l$ can be measured with $C_l^s>C_l^n$ is 
\be
     l^2_{\rm
     max}=\frac{\alpha(\nu,\Delta\nu)}{\beta(\nu,\Delta\nu)}
     f_{\rm cover}^2\, t_0.
\label{l_max}
\ee
Hence the dependence of this maximum scale on the observation
time and covering fraction of the experiment is relatively
simple, while the dependence on the frequency and bandwidth are
more elaborate.

In Ref.~\cite{KovKam} we constructed the minimum-variance
estimator,
\begin{eqnarray}
     \widehat{\kappa_{\vec L}} 
      =  -\frac{\Omega N_{\vec{L}}}{L^2 } 
      \sum_{\vec l} I_{\vec l} I_{\vec L
     -\vec l}\; \frac{\vec L \cdot(\vec l C_l + (\vec L-\vec l)
     C_{|\vec L-\vec l|})} { C_l^{\rm map} C_{|\vec
     L -\vec l|}^{\rm map}\Omega^2} 
     \label{eq:huestimator}
\end{eqnarray}
for the Fourier modes of the lensing convergence obtained from a
single redshift slice of 21-cm brightness temperature with
intensity Fourier coefficients $I_{\vec l}$ in a sky
patch of angular size $\Omega$.  Here $C^{\rm map}_l =  C_l^{\rm
n} + C_l$ is the sum of the signal and noise power spectra, and
\begin{equation}
     N_{\vec{L}}^{-1} = \frac{2\Omega}{L^4}      \sum_{\vec l} 
     \frac {
     [\vec L \cdot(\vec l C_l + (\vec L-\vec l)   C_{|\vec
     L-\vec l|})]^2}
     {C_l^{\rm map} C_{|\vec L -\vec l|}^{\rm map}
     \Omega^2},
\end{equation}
is the noise power spectrum for $\kappa_{\vec L}$. The variance with which $\kappa_{\vec L}$ can be measured is in turn given by  $\langle| \widehat{\kappa_{\vec L}} |^2\rangle=(2\pi)^2\delta(0)N_{\vec{L}}=\Omega N_{\vec{L}}$.

Under the assumption, Eq.~(\ref{l_max}), the noise in the limit
$L \ll l$ is approximately 
\be
     N_{\vec{L}} \sim \frac{4\pi}{
     l_\max^2}=\frac{4\pi\beta(\nu,\Delta\nu)}{\alpha(\nu,\Delta\nu)f_{\rm
     cover}^2\, t_0}.
\label{VarLimit}
\ee

Following Refs.~\cite{KovKam}-\cite{Book:2011dz}, we
can use a range of frequencies to cover a maximum number of
independent redshift slices given by
\be
     N_z\simeq 0.5\,l_{\rm max}(\Delta\nu/\nu)(1 + z)^{-1/2}.
\ee
This is roughly $N_z\sim1500$ for the frequencies corresponding
to the EoR at the maximum resolution of SKA.

Our goal will be to determine how well a parameter $p$ that
describes the lensing amplitude can be measured.  We shall choose $p$ to
be the fractional density contrast for the void and for the
texture it will be the deflection amplitude.  
Applying the estimator, Eq.~(\ref{eq:huestimator}), for the
convergence to a patch of sky around the cold spot, we can
retrieve a two-dimensional image of the weak-lensing convergence of the
structure. The total number of pixels in Fourier space is the same as in real space.  We then denote the $N$ Fourier wave numbers as $\vec {L_i}$ for $i=1,2,\ldots,N$. If
$\widehat{\kappa_{\vec{L}_i}}$ is the measured value for the pixel $i$ with variance
$\langle|\widehat{\kappa_{\vec{L}_i}}|^2\rangle=\Omega
N_{\vec{L}_i}$, and if the corresponding theoretical value is
$\kappa^{\rm th}_{L_i}(p_{\rm fid})$ (calculated for some
fiducial value for the desired parameter $p_{\rm fid}$), then
the estimator, along with its variance, is given by
\be
     \widehat{p}_i =
     \frac{\widehat{\kappa_{\vec{L}_i}}}{\kappa^{\rm
     th}_{\vec{L}_i}(p_{\rm fid})}p_{\rm fid},~~~~
     \langle|\widehat{p}_{i}|^2\rangle =  \frac{\Omega
     N_{\vec{L}_i}}{|\kappa^{\rm th}_{\vec{L}_i}(p_{\rm
     fid})|^2}p^2_{\rm fid}
\label{ParamPerPixel}
\ee
where the dependence of $\kappa$ on the parameter $p$ is linear (as is the case for the amplitudes of the weak lensing deflection angles, whose detection prospects we investigate here).
The minimum-variance estimator over the patch is then 
\be
\left. \widehat{p} =  \left(\sum\limits_i
\widehat{p}_i/\langle|\widehat{p}_{i}|^2\rangle\right)
\middle/ \left(\sum\limits_i
1/\langle|\widehat{p}_{i}|^2\rangle\right)\right. .
\ee
Its variance under the null hypothesis, for a spherically
symmetric real-space convergence profile, related to the deflection angle
by
$\kappa(\theta)=(1/2)\vec\nabla_\theta\cdot\boldsymbol\alpha(\theta)=(1/2\theta^2)
(\partial/\partial \theta)\left(\theta^2\alpha(\theta)\right) 
$, is given by \cite{KovKam}
\bea
     \sigma^{-2}_{p} &=&  \frac{N_z l_{\rm max}^2}{2p^2_{\rm
     fid}}\int\limits^\Lambda \, \theta\,
     d\theta \nonumber \\
     &  \times &  \left[
     \int\limits^\Lambda \, \varphi\, d\varphi\, W(\varphi)
     \int\limits_0^{2\pi} \, d\phi\,
     \kappa(\sqrt{\theta^2+\varphi^2+2\theta\varphi\cos\phi})
     \right]^2. \nonumber \\
\label{signaltonoise}
\eea
This equation, together with Eq.~(\ref{l_max}), provides the
signal-to-noise with which we can detect the properties of
either a void or a texture that might be responsible for the
WMAP cold spot for a chosen set of experimental parameters. 

We now consider a design for SKA based on an extended region of
$D\! \sim\! 6$ km which corresponds to a maximum angular
resolution of $\sim1\,{\rm arcmin}$ or $l_{\rm
cover}(\nu)\!\sim\! 10^4$ for the relevant frequencies of the
EoR with a coverage fraction  $f_{\rm cover}\! \sim\! 0.02$. We
assume that SKA will be able to cover the full frequency range
of the EoR in bandwidths of order $1\,{\rm MHz}$.
 
We first consider the prospects for detection of a void.  The
void explains the WMAP cold spot \cite{InoueSilk}
through the gravitational redshift imparted to a CMB photon by
the linear integrated Sachs-Wolfe (ISW) effect \cite{SachsWolfe}
and the second order Rees-Sciama effect \citep{ReesSciama} as
the photon passes through the void. Reference ~\cite{InoueSilk}
focused on a compensated  dust-filled void with a fractional
density contrast $\delta_V\! \sim\! -0.3$ centered at
$z\lesssim1$, and their analysis suggested that the comoving
radius of the region required to explain an observed anisotropy
$\Delta T/T\sim 10^{-5}$ (an underestimate of the WMAP cold spot
anisotropy) is $r\sim 200\mbox{-}300\, {\rm Mpc}/h$.  In Ref.~\cite{Rudnick}
an estimate based on the linear ISW effect alone for a
completely empty void ($\delta_V=-1$) at $z\lesssim 1$ led to an
estimated comoving radius of $r\sim120\, {\rm Mpc}/h$.

To simplify the treatment of weak lensing with the thin-lens
approximation, we consider a cylindrical void \cite{DasSpergel}
(which is justified as large voids often have large axis ratios
instead of a perfectly spherical shape) with its axis aligned
with the line of sight towards the WMAP cold spot. The comoving
radius $r$ of the cylinder determines its angular size
$R_V$ on the sky and with a comoving line-of-sight depth $L$ centered
at a redshift $z$, the cylinder can be approximated as a
disc of surface (under)density $\rho\delta_V
L/(1+z)$, where $\rho=3H_0^2/(8\pi G)$ is the critical
background density. The resulting expression for the deflection
angle $\boldsymbol\alpha_V=\alpha_V(\theta)\hat{\theta}$ is
given by \cite{DasSpergel}
\begin{equation}
\alpha_{V}(\theta) = 
\begin{cases}
A_{V} \theta, & \text{for  $\theta < R_{V}$},\\
 {A_{V}} \frac {R_{V}^2}  \theta, & \text{for $\theta\ge R_{V}$},
\end{cases}
\label{VoidDeflection}
\end{equation}
where the amplitude,
\ben
A_{V} =  \frac32 {\left(\frac{H_0}c\right)}^2
\left|\delta_V\right| \Omega_m L \frac{D_{LS}}{D_S} {D_L}{(1+ z)},
\een
depends on the comoving observer-lens ($D_L$), observer-source
($D_S$), and lens-source ($D_{LS}$) distances. To facilitate
comparison of our estimates for 21-cm lensing with CMB lensing,
we retain the parameter choices of Ref.~\cite{DasSpergel}: a
comoving radius $r=105\, {\rm Mpc}/h$ (yielding an angular
radius $R_V=3.1^{\circ}$ on the sky), a line-of-sight depth
$L=140\,{\rm Mpc}/h$, a redshift $z=0.8$, and a fractional
density contrast $\delta_V=-0.3$. The void considered here is not
compensated by a surrounding dense ring, but our cutoff
$\Lambda\sim R_V$ on the integration in
Eq.~(\ref{signaltonoise}) ensures that our signal-to-noise
estimates are unaffected by this. 

\begin{figure}[h!]
\includegraphics[width=0.8\linewidth]{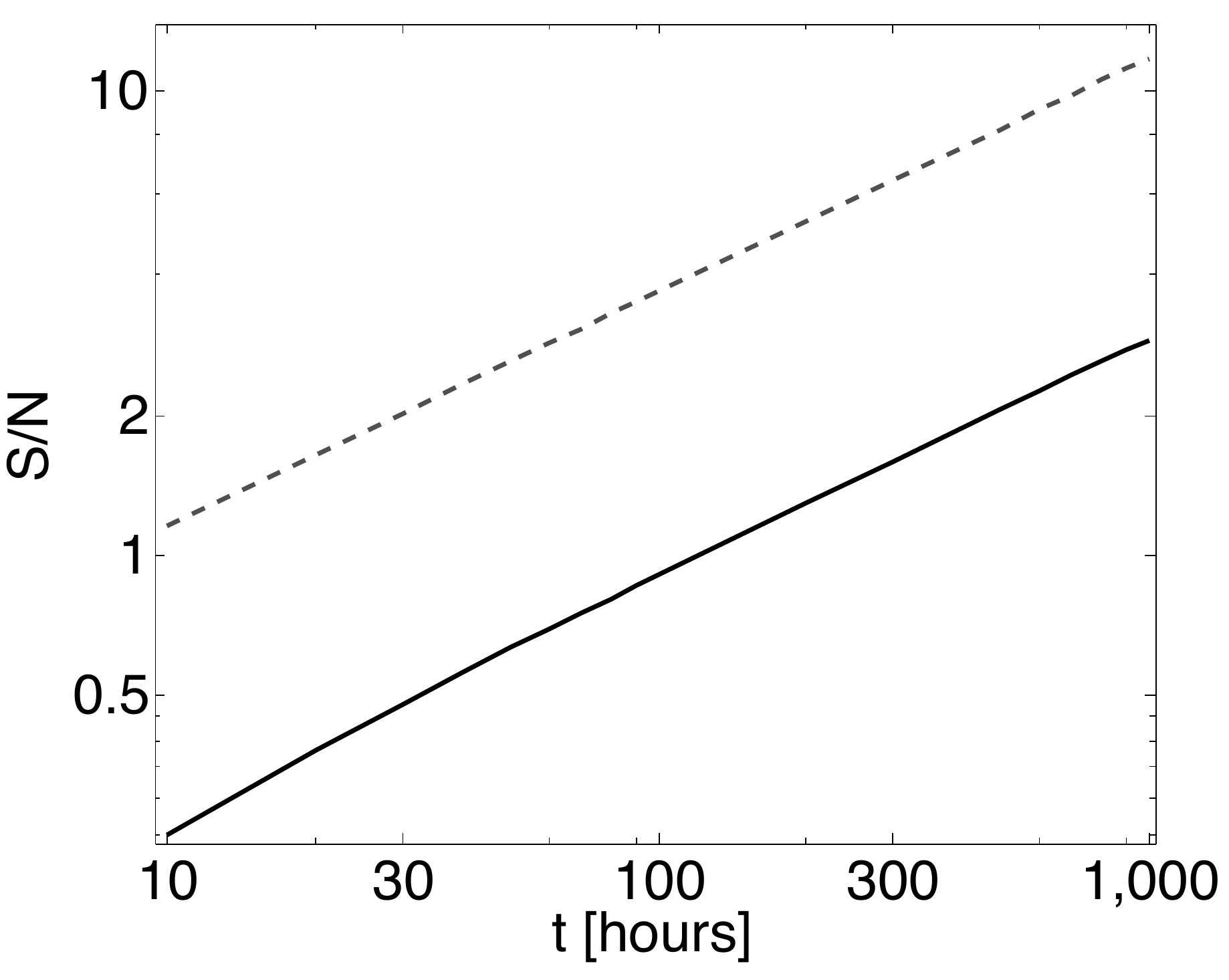}
\caption{The signal-to-noise with which SKA can detect a void
matching the WMAP cold spot as a function of observation time
over the field of view containing the cold spot, calculated numerically. In black we plot the prospects for the current plan of SKA and in dashed gray those for
SKA with four times the coverage fraction $f_{\rm cover}$.} 
\label{fig:SNVoid}
\end{figure}

Suppose that SKA observes a patch of sky containing the cold
spot with a bandwidth low enough to use the maximum number of
independent redshift slices in the EoR and enough exposure time
to reach its maximum angular resolution.  SKA would then be able
to detect such a void with a signal-to-noise ratio of
$\sim50\sigma$.  By contrast, the corresponding ratio for lensing
 of the CMB with an experiment with angular resolution comparable
to ACT or SPT but with negligible detector noise would be less than 
$2\sigma$. However, current plans to survey the complete
southern galactic sky with some frequency-dependent velocity
will allow limited observation time for each field of view. In
Fig.~\ref{fig:SNVoid}, we plot the signal to noise with which
such a void could be detected as a function of the dedicated
observation time in a patch of sky around the WMAP cold spot
with a bandwidth of $1\,{\rm MHz}$.

We now consider detection of a texture responsible for the WMAP
cold spot.  A cosmic texture is a topological defect composed of
localized, twisted configurations of fields produced during an
early-Universe phase transition that involves the breaking of a
symmetry of homotopy group $n=3$.  Unlike cosmic strings or
domain walls, which are stable once produced, the texture is
unstable and unwinds on progressively larger scales as the
Universe evolves.  The energy density associated with this
texture thus gives rise to a time-dependent gravitational
potential.  One consequence of such a potential is a cold or hot
spot in the corresponding direction in the CMB, depending on
whether the texture had collapsed before or after the observed
CMB photons crossed it. Another effect is the lensing deflection
of photons passing near this structure, which is our focus here.

A texture explanation for the WMAP cold spot 
\cite{ScienceTexture} requires a texture at redshift $z=6$
with a characteristic scale parameter $R_{T} \sim 5^{\circ}$, and
a bias-corrected symmetry-breaking scale $\epsilon\sim 4\times
10^{-5}$.  Lensing of photons by such a
spherically symmetric texture leads to a deflection angle
\cite{DasSpergel},
\ben
     \alpha_{T}(\theta) = A_{T}\theta  \left[1+ 4
     (\theta/R_{T})^{2} \right]^{-1/2},
\label{TextureDeflection}
\een
where $A_{T} =   (2\sqrt{2} \epsilon/R_{T}) (D_{LS}/D_{S})$ is
the deflection amplitude of the texture.

The texture will be, unlike the void, very hard to detect. One
reason is that its deflection angle is more than an order of
magnitude smaller than that of a void. This is because the 
characteristic time scale for the change of the void's  gravitational potential, 
the light crossing time, is much smaller than that of the void (roughly the Hubble time).  
The texture thus requires a much smaller energy-density perturbation 
than the void to explain the temperature decrement at the cold spot  \cite{DasSpergel}.
The other limiting factor is its relatively high redshift, $z=6$, which is near the
redshifts of the EoR.  Even with the optimal angular resolution
of SKA, a cosmic texture with parameters that fit the WMAP
cold spot will be only marginally detectable, with $S/N\gtrsim3$, 
requiring multiple redshift slices at frequencies corresponding to redshifts 
$z\gtrsim10$. Relaxing the redshift constraint down to $z\!=\!4$, which corresponds 
to the $95\%$ C.L. limit found in Ref.~\cite{ScienceTexture}, and using the full redshift
volume of the EoR, would yield an ideal bound of $S/N\gtrsim5$.

The scenario considered in Ref.~\cite{ScienceTexture} predicts
an abundance of $O(100)$ smaller ($\gtrsim1^{\circ}$) cold
texture spots in the CMB, providing another way to test the
texture hypothesis. A recent template-based bayesian search for
textures in WMAP CMB data \cite{Feeney} placed a limit of
$\sim5$ on the number of textures with the same symmetry
breaking scale as Ref.~\cite{ScienceTexture}, but did identify a
couple of best candidates (including the cold spot itself). With
a superior 21-cm lensing experiment, it might be possible to use
lensing reconstruction to study promising texture candidates
with high accuracy.

We have examined the ability of SKA to detect the large void or a cosmic texture that have been hypothesized to explain the WMAP cold spot by seeking the lensing distortion they induce in maps of 21-cm fluctuations from the EoR.
While a void would be easily detectable with SKA
at its ideal resolution (with unlimited observation time) and
still detectable even with a more realistic allocation of
observation time, a texture responsible for the WMAP cold spot
would most likely remain undetectable by an SKA-like experiment. 

We note that the effects of non-Gaussianities in the 21-cm radiation caused by nonlinear structure \cite{LuPen,LuPenDore}, which would induce a connected four-point contribution to the lensing estimator, which are relevant during the epoch of reionization (when certain patches of the IGM become substantially ionized well before its end), have been neglected here, as we have focused on the measurement of local distortions to the two-point correlations in the vicinity of a specific CMB feature. In addition, under the assumption that non-linearities from patchy reionization appear in scales below several ${\rm Mpc}/h$ during the EoR (corresponding to resolutions higher than those of SKA), there would be no contribution to our estimator.

Plans for a futuristic experiment \cite{Lunar}, based on placing a dark ages
observatory on the far side of the Moon, provide more promising prospects for these detections. A lunar experiment with a baseline on the order of $\gtrsim10\mbox{-}100\,{\rm km}$ would yield a corresponding angular resolution of $l_{\rm max}\sim10^4\mbox{-}10^5$ for source redshifts $z=30\mbox{-}300$. With a wide frequency range to cover a significant portion of the dark ages redshift volume and enough observation time to compensate for the covering fraction and high system temperature, both structures considered here would be easily detected.

It remains to be seen if indeed the lensing of 21-cm
fluctuations will provide a clue as to the nature of the cold
spot. A detection of either one of the local lensing sources
considered here would be of great importance as the large voids
required are extremely unlikely in $\Lambda {\rm CDM}$ and the
detection of textures would have important implications for the
study of high-energy theories.  Along
these lines one could also consider the detectability of other
models that might be related to CMB anomalies
\cite{Fialitzkov,Rathaus,GiantRings}.

E.D.K was supported by the National Science Foundation under Grant No. PHY-0969020 and by the Texas Cosmology Center. This work was supported at Johns Hopkins by DoE SC-0008108 and Grant No. NASA NNX12AE86G.

\end{document}